\documentclass{aastex62}

\usepackage{graphicx}

\def\sax{SAX~J1748.9-2021\,}

\graphicspath{{./}{figures/}}

\shorttitle{Differential reddening in NGC~6440}
\shortauthors{Pallanca et al.}

\begin{document}

\title{ High-resolution extinction map in the direction of the bulge globular cluster NGC~6440\footnote{Based on observations
    collected with the NASA/ESA HST (Prop. 11685, 12517, 13410, and 15403 ), obtained at the
    Space Telescope Science Institute, which is operated by AURA,
    Inc., under NASA contract NAS5-26555.}}

\correspondingauthor{Cristina Pallanca}
\email{cristina.pallanca3@unibo.it}
\author[0000-0002-7104-2107]{Cristina Pallanca} \affil{Dipartimento di Fisica e Astronomia, Universit\`a di Bologna, Via Gobetti 93/2, Bologna I-40129, Italy}  \affil{Istituto Nazionale di Astrofisica (INAF), Osservatorio di Astrofisica e Scienza dello Spazio di Bologna, Via Gobetti 93/3, Bologna I-40129, Italy}
\author[0000-0002-2165-8528]{Francesco R. Ferraro} \affil{Dipartimento di Fisica e Astronomia, Universit\`a di Bologna, Via Gobetti 93/2, Bologna I-40129, Italy}  \affil{Istituto Nazionale di Astrofisica (INAF), Osservatorio di Astrofisica e Scienza dello Spazio di Bologna, Via Gobetti 93/3, Bologna I-40129, Italy}
\author[0000-0001-5613-4938]{Barbara Lanzoni} \affil{Dipartimento di Fisica e Astronomia, Universit\`a di Bologna, Via Gobetti 93/2, Bologna I-40129, Italy}  \affil{Istituto Nazionale di Astrofisica (INAF), Osservatorio di Astrofisica e Scienza dello Spazio di Bologna, Via Gobetti 93/3, Bologna I-40129, Italy}
\author[0000-0003-4746-6003]{Sara Saracino} \affil{ Astrophysics Research Institute, Liverpool John Moores University, 146 Brownlow Hill, Liverpool L3 5RF, UK}\affil{Istituto Nazionale di Astrofisica (INAF), Osservatorio di Astrofisica e Scienza dello Spazio di Bologna, Via Gobetti 93/3, Bologna I-40129, Italy}
\author[0000-0003-4592-1236]{Silvia Raso} \affil{Dipartimento di Fisica e Astronomia, Universit\`a di Bologna, Via Gobetti 93/2, Bologna I-40129, Italy} \affil{Istituto Nazionale di Astrofisica (INAF), Osservatorio di Astrofisica e Scienza dello Spazio di Bologna, Via Gobetti 93/3, Bologna I-40129, Italy}
\author[0000-0002-4282-3801]{Paola Focardi}\affil{Dipartimento di Fisica e Astronomia, Universit\`a di Bologna, Via Gobetti 93/2, Bologna I-40129, Italy}

\begin{abstract}
We used optical images acquired with the UVIS channel of the Wide Field Camera 3 on board of the  {\it Hubble Space Telescope} to construct the first  high-resolution extinction map in the direction of NGC 6440, a globular cluster located in the bulge of our Galaxy. 
The map has a spatial resolution of $0.5 \arcsec$ over a rectangular region of about $160 \arcsec \times 240 \arcsec$ around the cluster center, with the long side in the  North-West/South-East direction.  We found that the absorption clouds show patchy and filamentary sub-structures with extinction variations as large as $\delta {\rm E}(B-V)\sim0.5$ mag.  We also performed a first-order proper motion analysis to distinguish cluster members from field interlopers. After the field decontamination and the differential reddening correction, the cluster sequences in the color-magnitude diagram appear  much better defined, providing the best optical color-magnitude diagram so far available for this cluster.
\end{abstract}


\section{INTRODUCTION}\label{intro}
As part of the project  {\it Cosmic-Lab}, we are conducting a systematic study of the kinematical properties and dynamical status of a sample of Galactic globular clusters (GCs, see \citealt{ferraro12,ferraro18a,ferraro18b, lanzoni16,lanzoni18a,lanzoni18b, miocchi13}) harbouring a variety of exotic  stellar populations \citep{bailyn95}, like interacting binaries \citep{pooley03}, blue stragglers \citep{ferraro09a,ferraro16a,beccari19}, and millisecond pulsars (MSPs, \citealt{ransom05}). 
Concerning to the latter subject we are carrying on an extensive search for  optical counterparts to MSPs in different stages of  their formation and evolutionary path (see \citealt{ferraro01,ferraro03,ferraro15, pallanca10,pallanca13,pallanca14, cadelano15a,cadelano15b,cadelanoM3}). 
In this respect the case of the GC NGC 6440 is particularly intriguing since  it hosts six (classic) radio  MSPs \citep{freire08} and an accreting MSP \citep[\sax, ][]{AMSP6440}.  

NGC 6440 is a metal-rich GC ([Fe/H]$\sim-0.56$, \citealt{origlia97, origlia08a}) located in the Milky Way bulge, 1.3 kpc away from the center of the Galaxy  \citep{harris}. 
Given its mid distance  from the Sun and relatively high reddening  \citep[d=8.5 kpc and $ {\rm E}(B-V)=1.15$, ][]{valenti04,valenti07}, it has been  poorly  investigated from the photometric point of view.
To date, only a few optical color-magnitude diagrams  \citep[CMDs; ][]{piottoSNAP, ortolani94} and a few studies in the  infrared \citep{valenti04,origlia08a, mauro12, minniti10} are  available. 
These studies confirmed a large value of the reddening in the cluster direction, due to its location inside the Galactic bulge and the presence of a strong and complex differential reddening \citep{munoz17}. 
Interestingly, NGC 6440 was one of the clusters indicated by \citet{mauro12} to show in the CMD a possible split in the HB. 
This is particularly intriguing since a similar feature, combined with the discovery of multi-iron sub-populations \citep{ferraro09b,origlia13,massari14}, brought to classify Terzan 5 as a complex stellar system, possibly the fossil of a primordial fragment of the bulge formation epoch (see \citealt{ferraro16b,lanzoni10,origlia19}) after a 40 years long conviction that it was just a common GC.     

Recently, in the context of the {\it Cosmic-Lab} project our group secured an ultra-deep set of high resolution images obtained with the  Wide Field Camera 3 (WFC3) on-board the {\it Hubble Space Telescope} ({\it HST}) to explore the innermost regions of the cluster. 
These observations were used to identify the optical counterpart to the accreting MSP \sax  (see  \citealt{cadelano6440}). 
Here we use this dataset, complemented with archive observations, to construct  the first high resolution differential reddening map of the innermost regions of the cluster\footnote{A free tool providing the color excess values at any coordinate within the investigated Field of View can be found at the Web site http://www.cosmic-lab.eu/Cosmic-Lab/Products.html.}.

The so-called interstellar reddening is a phenomenon that alters the properties of the electromagnetic radiation emitted by a source  and it is due to the absorption and scattering of the radiation (preferentially at short wavelengths)  produced by dust clouds along the light pathway. 
 As a result a star appears systematically fainter and redder than its effective temperature would imply (from this the word reddening).
 The entity of the reddening is usually parametrised by the color excess  $ {\rm E}(B-V)$ defined as the difference between the observed color $(B-V)$ and the intrinsic color $(B-V)_0$. Once this quantity is defined, the absorption coefficient depends on  the wavelength, significantly increasing toward shorter  wavelengths \citep{cardelli},
 and on the target effective temperature and metallicity \citep{casagrande,ortolani17,kerber}.
Spatially variable extinction, or differential reddening, occurs in all directions throughout the Galaxy and is also present across the field of view of most GCs. 
This induces a systematic broadening of the evolutionary sequences in the CMD, thus hampering the accurate 
 characterization of their  photometric properties, the identification of multiple sub-populations, and, in general, preventing a precise determination of fundamental GC parameters \citep[as, for instance, the age; e.g., ][]{bonatto13}.

So far, in the literature several methods  have been proposed to model and correct for the differential spatial variations of reddening.
Basically, two main approaches can be found. 
 The first one, which we may  name the ``cell by cell'' approach, consists in dividing the observed region of the GC in cells of constant dimension, and to calculate the differential reddening value for each cell \citep{heitsch99,piotto99,vonbrau01,mcwilliam10,nataf10,gonzalez1,gonzalez2,massariTer5,bonatto13}. The second one, which can be  named the ``star by star'' approach, consists in estimating the reddening of each star on the basis of the spatially closest objects \citep{milone12,bellini13,saracino6569}.
Beyond the method, also the reference objects used  to  estimate  the differential reddening may be different:  main sequence (MS), horizontal branch (HB), red giant branch (RGB) stars, and even variable RR Lyrae \citep[][and references therein]{alonco11}. 
In all cases, in order to maximise the spatial resolution of the derived differential reddening map, the displacement of as many stars as possible should be studied. For this reason, here we adopt a ``star by star''  approach based on the displacement of faint RGB, sub giant branch (SGB) and bright MS cluster stars with respect to the cluster fiducial 
mean ridge line of these sequences in the CMD,  taking into account the dependence of the absorption coefficient both on the wavelength and on the effective temperature of the targets. The paper is organized as follows. In Section 2 we summarize the dataset and the main steps of the photometric analysis. Section 3 is dedicated to a description of the proper motion analysis performed to distinguish cluster members from Galactic field interlopers. 
In Section 4 we present the method used in this paper to estimate the differential reddening of the cluster. In section 5 we discuss the main results of the paper and summarize the conclusions.

\section{OBSERVATIONS AND DATA ANALYSIS}
\subsection{Dataset}
\begin{figure*} 
\begin{center}
\includegraphics[width=90mm]{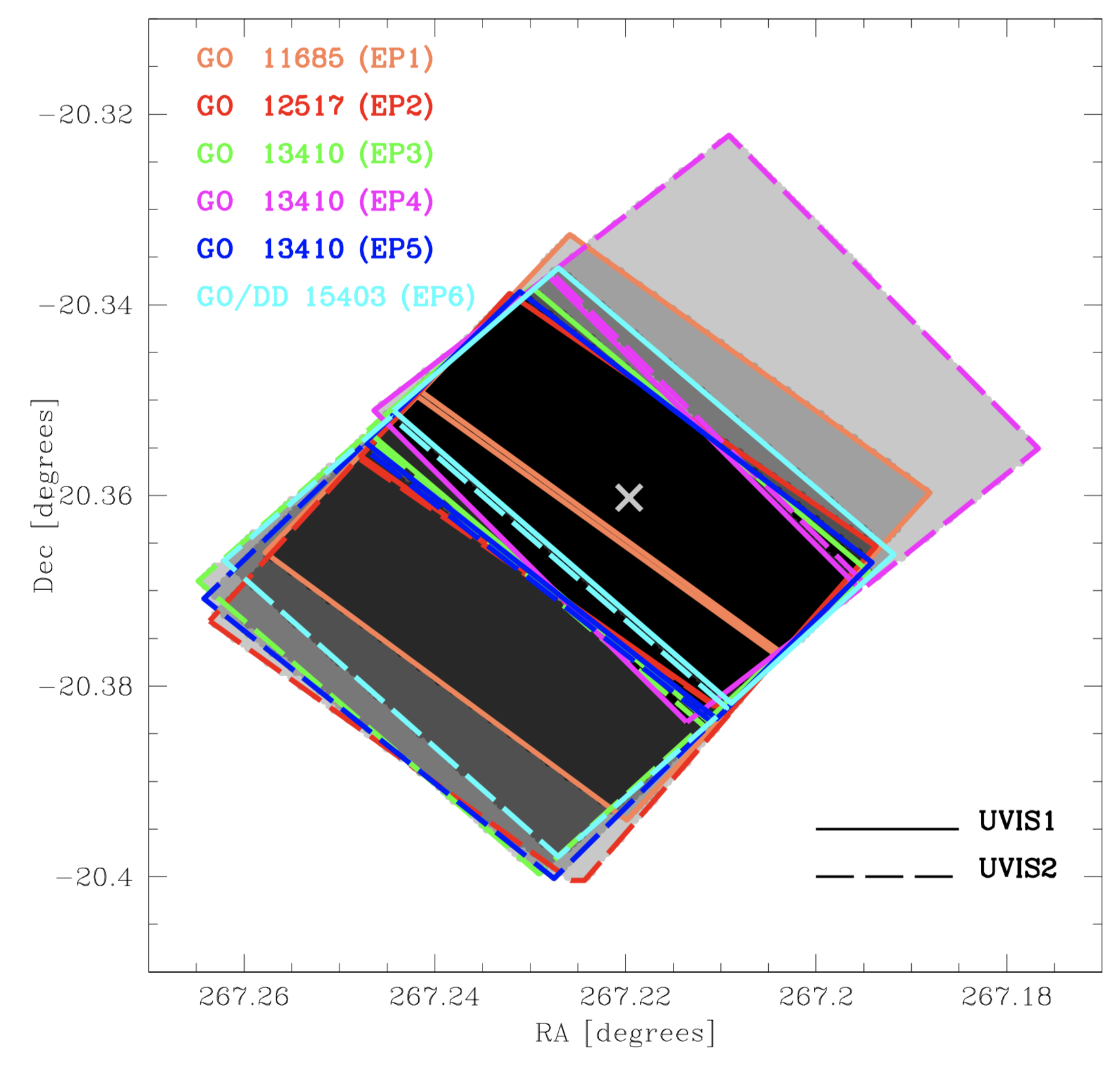}
  \caption{Total field of view  sampled by the observations. The cross marks the cluster center \citep[][2010 version]{harris}. Solid and dashed lines identify the UVIS1 and UVIS2 chips of the WFC3, respectively, while different colors represent different epochs (see the legend in Figure). Different regions are shaded with increasingly dark grey colors  as a function of increasing number of images  overlapped in those portions of the FOV. These range from only one in the case of most of UVIS2 of GO 13410 (EP4) up to all the six epochs at the cluster center }
\label{Fig:FOV}
\end{center}
\end{figure*}

The photometric dataset used for this work consists of a large set of {\it HST} high-resolution images obtained with the ultraviolet-visible (UVIS) channel of the WFC3. 
Most of  these images have been acquired under a few programmes aimed at studying the MSP content of the cluster (GO 12517, PI: Ferraro; GO 13410 and GO/DD 15403, PI: Pallanca). A few additional images  (GO 11685, PI:Van Kerkwijk) have been taken from the Archive. In summary, a total of 126 long exposure images, acquired under 4 different programs, has been analyzed.  The details of the observations are reported in Table \ref{Tab:dataset}.
All the images are dithered by few pixels, in order to avoid spurious effects due to bad pixels. 
The images were acquired in six different epochs (EP1-6), with different pointings (e.g., the cluster center in EP1 is located onto the UVIS aperture, while for EP2/3/4/5 is onto the UVIS1) and different telescope position angle (e.g EP4 is rotated by  $\sim 180$ degrees with respect to all other epochs). 
This makes the   total field of view (FOV)  of the observations used in this work larger than the  WFC3 nominal FOV  ($\sim160\arcsec \times160\arcsec$ arcsec),  covering a rectangular region of about $160 \arcsec \times 240 \arcsec$ roughly centred on the cluster center and with the long side in the  North-West/South-East  direction (see Figure \ref{Fig:FOV}).

\begin{table}
\begin{center}
\begin{tabular}{l|l|c|c|c|c|l}
\hline
  Epoch & Program ID & PI & MJD & YEARS & Filter & $\rm N_{exp} \times \rm T_{exp}$\\
\hline
\hline
EP1 & GO 11685 & Van Kerkwijk & $55052$ & 2009.60 &F606W & 1$\times$392 s + 1$\times$348 s   \\ 
       &           &              &            & & F814W & 1$\times$348 s + 1$\times$261 s   \\ 
\hline
EP2 & GO 12517 & Ferraro & $56128$&  2012.13 &F606W & 27$\times$392 s   \\
         &         &              &             & &F814W & 27$\times$348 s   \\
\hline
EP3 & GO 13410 & Pallanca & $56585$ & 2013.80&F606W & 5$\times$382 s   \\
        &          &              &             & & F814W & 5$\times$222 s \\
        &          &              &             & & F656N & 10$\times$934 s   \\
EP4 &        &              & $56798$ & 2014.38 &F606W & 5$\times$382 s   \\
        &          &              &             & & F814W & 5$\times$222 s   \\
        &          &              &             & & F656N & 10$\times$934 s   \\
EP5 &        &              & $56908$ & 2014.68 &F606W & 5$\times$382 s     \\
        &          &              &             & & F814W & 4$\times$222 s + 1$\times$221 s  \\ 
        &          &              &             & & F656N & 6$\times$934 s + 2$\times$864 s +2$\times$860 s  \\
\hline
EP6 & GO/DD 15403 & Pallanca & $58056$ & 2017.83&F606W & 2$\times$382 s   \\
        &          &              &             & &F814W & 1$\times$223 s  + 1$\times$222 s  \\
        &          &              &             & & F656N & 2$\times$969 s + 2$\times$914 s    \\                 
\hline
\end{tabular}
\end{center}
\caption{Summary of the used dataset. The first five columns report the details of the observations (the labels used along the paper, program ID, principal investigators,  approximated Modified Julian Days and the years of the visits), while the last two list the filters and the exposure times of each single image.}
\label{Tab:dataset}
\end{table}

\subsection{Photometric analysis, astrometry and magnitude calibration}\label{Sec:red}
The data reduction procedure has been performed independently on each different epoch (see Table \ref{Tab:dataset}) and detector (UVIS1 and UVIS2) onto the CTE-corrected (\texttt{flc}) images after a correction for Pixel-Area-Map by using standard \texttt {IRAF} procedures.
The photometric analysis has been carried out by using the \texttt{DAOPHOT} package \citep[][]{daophot}. 
For each image we first modelled the  point spread  function (PSF) by using a large number ($\sim 200 $) of bright and nearly isolated stars.
The adopted PSF consists of an analytical \texttt{MOFFAT} model plus a second order spatially variable look up table.
We then performed a source search onto all single images imposing a $5 \sigma$ threshold to the F606W and F814W images and $10 \sigma$ to  F656N images. 
We used such a conservative threshold in order to have a catalog free from spurious objects, while keeping  a relatively high number of sources.

After applying the PSF previously modelled  to each single image, we constructed a master list for each combination of epoch and detector,
 considering as reliable sources all the objects measured in more than half of the images in at least one filter. 
We then run the \texttt{ALLFRAME} package \citep[][]{daophot,allframe} onto all the images of each single group ``epoch/detector'' by using the master list above. 
The final catalogs (one per each epoch and detector) contain all the objects measured in at least 2 filters. 
For each star, they list the instrumental coordinates, the mean magnitude in each filter and two quality parameters ({\it chi} and {\it sharpness})\footnote{The {\it chi} parameter is the ratio between the observed and the expected pixel-to-pixel scatter between the model and the profile image. The {\it sharpness} parameter quantifies how much the source is spatially different from the PSF model. In particular, positive values are typical of more extended sources, as galaxies and blends, while negative values are expected in the case of cosmic rays and bad pixels. }. 

The instrumental magnitudes of each ``epoch/detector'' catalog have been independently calibrated to the VEGAMAG system by using  the photometric zero points and the procedures reported on the WFC3 web page\footnote{http://www.stsci.edu/hst/wfc3/phot\_zp\_lbn}.
In order to guarantee the photometric homogeneity of the catalogs, we used the large number of stars in common among different catalogs to quantify and correct any residual photometric offsets. 
To this aim, we adopted the EP2/UVIS1 catalog as reference since this pointing has the largest photometric accuracy, thanks to its largest number of exposures (see Table \ref{Tab:dataset}), and it samples the cluster center.  Very small residuals (smaller than 0.05 mag) have been measured and applied to the photometric catalogs to match the reference. 

It is well known that WFC3 images are affected by geometric distortions within the FOV, hence we corrected the instrumental positions of the stars in each catalog by applying the equations and the lookup table reported by \citet[][]{distortionWFC3}. 

Finally, since a non negligible contamination from field stars is expected, the distortion-corrected catalogs have been placed onto the  International Celestial Reference System (ICRS)  \citep[GAIA DR2;][]{gaia16a,gaia16b,gaiadr2}  by using only a sub-sample of stars with large probability to be cluster members. 
To this end, we performed a first-order estimate of the stellar proper motions, as described in  the next Section.

\section{ RELATIVE PROPER MOTIONS}\label{Sec: pm}

NGC 6440 is a GC located in the Milky  Way bulge, hence a strong contamination by field stars (both of disk and bulge) is expected.
Given the  procedure adopted in this work to estimate the differential reddening (see Section \ref{Sec:reddening}), a separation of the cluster population from the field  is useful to better constrain the extinction. 
Unfortunately, the comparison between a Galaxy simulation in the direction of NGC 6440 \citep{besancon1,besancon2} and the cluster absolute proper motion measure \citep{gaiaPM} showed that  NGC 6440 is moving in the same direction of all the Milky Way populations (both disk and bulge) on the plane of the sky,  making   decontamination of the sample from field star interlopers  quite complex in this case. 

While the detailed analysis of proper motions will be  presented in a forthcoming paper (Pallanca et al., 2019, in preparation), here we  illustrate just the main steps of the procedure  that we adopted to obtain a first-order separation of cluster members from field stars.

\begin{figure*}
\begin{center}
\includegraphics[width=90mm]{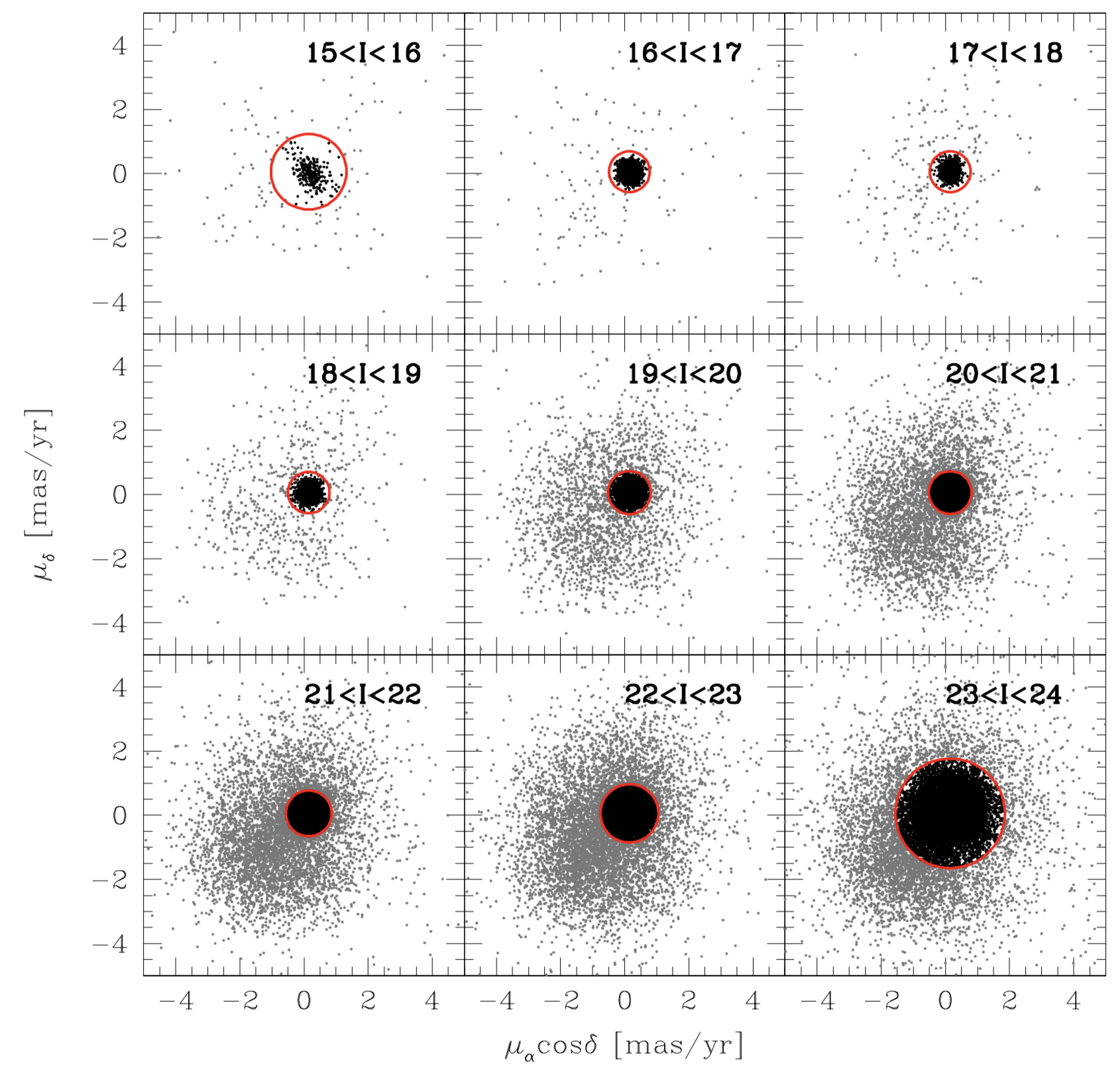}
  \caption{Vector point diagram for the magnitude bins labelled in each panel. 
  The red circles mark the adopted radii (see text for details on its determination) to separate cluster stars (black dots) from field stars (gray dots). Clearly, the two populations are overlapped in the VPD, hence a complete decontamination is impossible.}
\label{Fig:vpd}
\end{center}
\end{figure*}

Table \ref{Tab:dataset} shows that our sample contains a large number of repeated exposures with a relatively large acquisition time separation (up to 8 yrs).   
 In principle, we could use only the pair of epoch separated by the largest temporal baseline (i.e., EP1 and EP6), in order to obtain a first-order separation of cluster and field stars.  
However, these two epochs are those with the smallest number of acquired exposures (only two per filters; see Table \ref {Tab:dataset}), hence they are not ideal to get high photometric accuracy and consequently accurate centroid positions (which are crucial to derive high-quality proper motion measures). 
Thus, we decided to  consider  all the epochs in order to maximise both the number of measurements in as many epochs as possible,  and the size of the sampled FOV.
Moreover,  to minimise the effect of possible field contamination in the sample of stars used to compute the coordinate transformation, we adopted a two-step procedure.
In the first step we indeed  combined EP1 and EP6 only (i.e., the most distant in terms of time) to build a zero-order vector point diagram (VPD),  which we used to make a  pre-selection of bona fide member stars  distributed along the  entire FOV, by excluding the most evident field interlopers.
We then matched\footnote{We used CataXcorr,  a code aimed at cross-correlating catalogs and finding astrometric solutions, developed by P. Montegriffo at INAF - OAS Bologna. This package has been successfully used in a large number of papers of our group in the past years.} the list of these candidate members  with the VVV survey \citep[][]{VVVa,VVVb}  sample in the direction of NGC 6440, after  placing the latter on the ICRS astrometric system through cross-correlation with 
 Gaia DR2 data, used as astrometric reference.
In  the second step we reported all the single epoch positions to    the reference system of the bona fide member star catalog  and we calculated the final proper motion taking into account (for each star) all the centroid  measurements from all the epochs. 
To this aim, for each star we fitted all the known positions with a $\chi ^2$ method according to the following equations:
\begin{equation}
RA_{\rm obs}=\Delta RA \cdot (MJD_{\rm obs}-MJD_{\rm 2000})+RA_{\rm 2000}
\end{equation}
\begin{equation}
Dec_{\rm obs}=\Delta Dec \cdot (MJD_{\rm obs}-MJD_{\rm 2000})+Dec_{\rm 2000}
\end{equation}
 where $RA_{\rm obs}$, $Dec_{\rm obs}$ and $MJD_{\rm obs}$ are the coordinates  and the modified Julian day of  each single observation, while $RA_{\rm 2000}$ and $Dec_{\rm 2000}$ are the positions at the reference epoch J2000.
The final relative proper motions are $\mu_\alpha=14600\cdot\Delta RA$ and $\mu\delta=14600\cdot\Delta Dec$ expressed in mas/yr.
The final VPD is reported in Figure \ref{Fig:vpd}. 

\begin{figure*}
\begin{center}
\includegraphics[width=180mm]{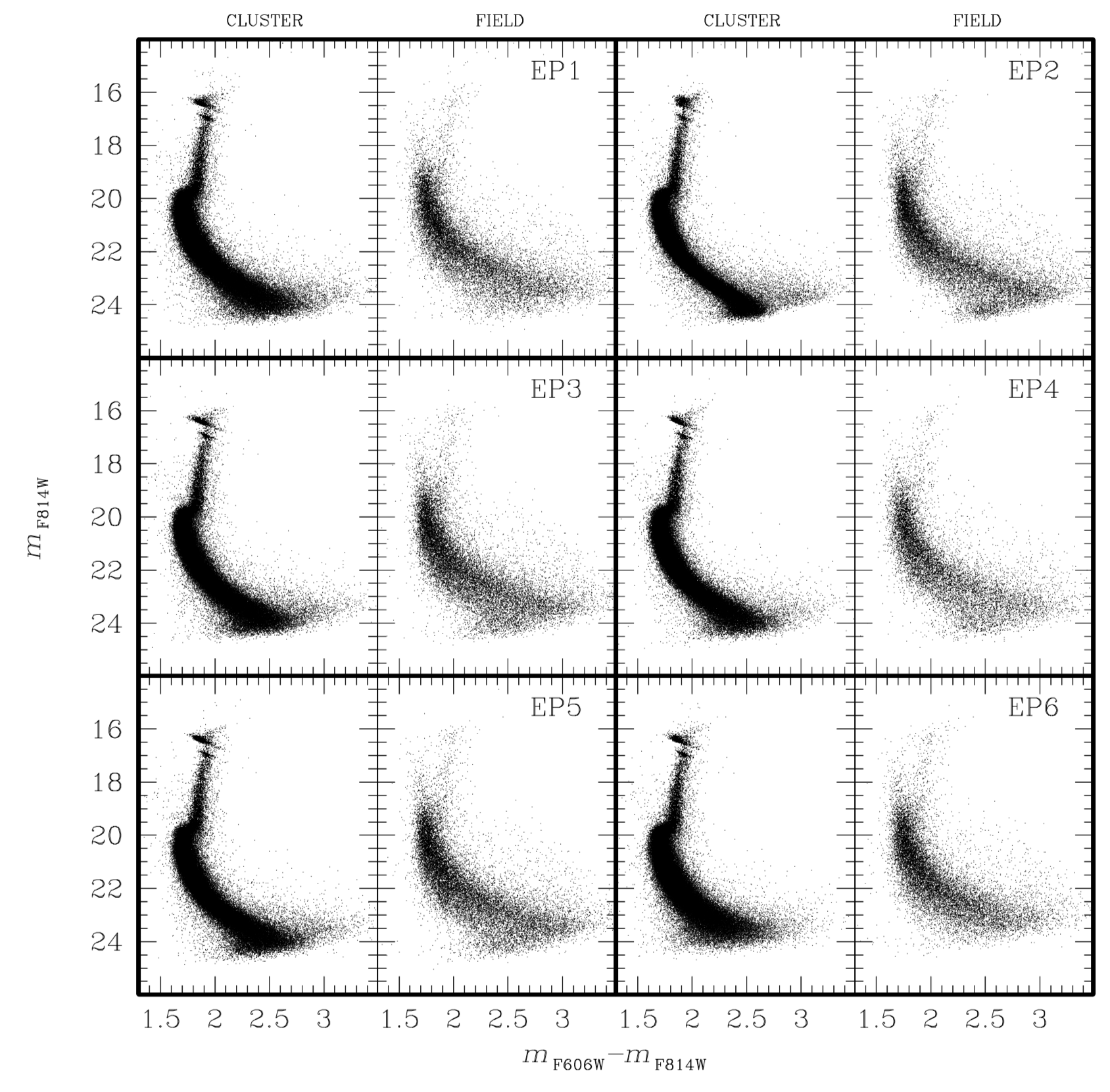}
  \caption{ For each epoch (see labels) two CMDs are shown: in the left panel we plotted all the candidate member stars, while in the right panel we show the  non member stars, according to the selection shown in Figure \ref{Fig:vpd}.}
\label{Fig:cmdPM}
\end{center}
\end{figure*}

As expected, the cluster stars appear as a bulk of points with a highly peaked proper motion distribution around the origin of the VPD, while field stars follow a much more dispersed  distribution. 
Although the significant overlap between the two distributions in the VPD  prevents a complete decontamination from the field interlopers,  the significantly different shape of the two distributions surely allows to perform a first-order identification of likely field interlopers.
To this end, we estimated the   dispersion ($\sigma_{PM}$) of the proper motion distribution around the origin of the VPD in each magnitude bin and assumed   the $2.5\sigma_{PM}$ radius as a  “cluster confidence” radius ($r_{cc}$), able to separate the portion of the VPD dominated by cluster stars ($r<r_{cc}$) from that dominated by field interlopers ($r>r_{cc}$). 
In the following, according to what  is shown in  Figure \ref{Fig:vpd}, we consider cluster members  those stars lying in the VPD at $r<r_{cc}$, being aware that  a residual field contamination is present. 
The resulting CMDs, obtained from the selected cluster members and field interlopers separately, are shown in Figure \ref{Fig:cmdPM} for each epoch.

\section{DIFFERENTIAL REDDENING}\label{Sec:reddening}
\begin{figure}
\begin{center}
\includegraphics[width=90mm]{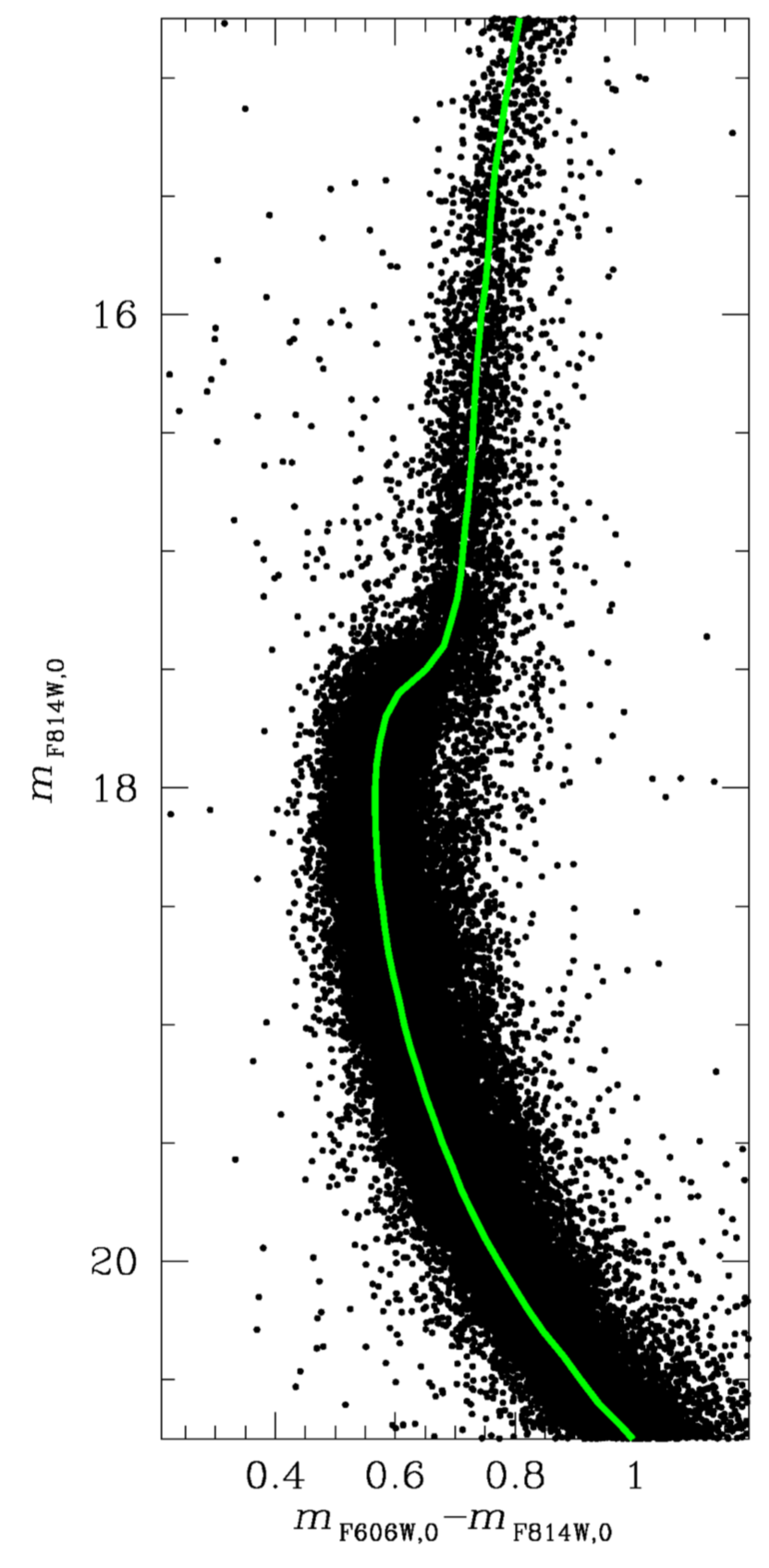}
  \caption{ Dereddened cluster reference mean-ridge line (D-MRL, in green) obtained from the plotted sub-sample of stars (black dots) selected as described in Section \ref{Sec:reddening}.}
\label{Fig:lmed}
\end{center}
\end{figure}

 The aim of this work is to derive  a differential extinction map in the direction of the globular cluster NGC 6440, at the highest spatial resolution achievable with the available data.  
To do this, we adopted a procedure already tested in previous works (see, e.g., \citealt{milone12,bellini13,saracino6569}),  also taking into account the dependence of the extinction coefficients on the effective temperature of the stars.
As discussed in the Introduction,  to maximise the spatial resolution here we adopted the ``star by star'' approach and used the sources observed along the RGB, SGB and bright MS.
 As a preliminary step, we assigned a temperature to each star, using a DSED isochrone model\footnote{We adopted a 13 Gyr isochrone with [Fe/H]=-0.56, [$\alpha$/Fe]=0.4, ${\rm E(B-V)=1.18}$ and $(m-M)_0=14.3$.} \citep{dotter}. 
We then dereddened the observed magnitudes according to the relations of \citet{casagrande}. On this preliminary dereddened diagram ($m_{\rm F606W,0}-m_{\rm F814W,0}$,  $m_{\rm F814W,0}$), we computed the dereddened  mean ridge line (D-MRL).
To this end, we divided the considered portion of the CMD in different magnitude bins, and in each bin we computed the mean color after a $3\sigma$-clipping rejection. 
We adopted variable mag-wide bins (ranging from 0.2 to 0.5 mag) in order to best sample the morphology changes of the sequences in the MS Turn-Off region. 
In order to compute the D-MRL, we used only stars in the “reference” catalog (EP2/UVIS1), sampling the cluster core with the highest photometric quality (see Section \ref{Sec:red}).  
Moreover, only  cluster member candidates (as selected  from the relative proper motions as described in the previous section) have been considered. 
This sample was further cleaned by applying a 3-sigma rejection on the {\it chi} and {\it sharpness} parameters. 
The D-MRL was further refined onto the first-step differential reddening corrected catalog and then applied to all the photometric catalogs. 
The final D-MRL is shown (as a green line) in Figure \ref{Fig:lmed} .

The procedure then consists in determining,  per each star in the catalog, the value of ${\rm E}(B-V)$ required to shift the D-MRL line along the direction of the reddening vector to fit the CMD drawn by $N_*$ sources spatially lying in the  neighborhood of that star. 
The ideal number of stars ($N$) for  sampling  the CMD  would be a few dozens.
However, to keep the spatial resolution as  high as possible, we introduced a complementary input parameter defined as the limiting radius ($r_{\rm lim}$) of a circle centred on each star, including the sources used to build the CMD.  
We set $N=30$ and $r_{\rm  lim}= 10\arcsec$. 
As soon as one of these two conditions is satisfied, the other is  not considered anymore.  
Typically, the number of stars ($N_*$) within a $10\arcsec$ radius is much larger than $N=30$ in the inner cluster regions, while the opposite is true in the outskirts. 
Hence, the typical input parameter used at small radii is $N$, while it becomes $r_{\rm lim}$ at large cluster-centric distances.

So, following either the $N$ or the $r_{\rm lim}$ criterion, for each star of the catalog we selected from the bona fide star list the $N_*$ closest objects in the magnitude range $17.5<m_{\rm F814W}<21.5$. 
We then shifted the D-MRL along the reddening direction in steps of ${\rm E}(B-V)$ by  adopting the absorption coefficients calculated as a function of the wavelengths and effective temperature of the targets  \citep{casagrande}.
For each ${\rm E}(B-V)$ step, we calculated the residual color $\Delta VI$ defined as the difference between the $(m_{\rm F606W}-m_{\rm F814W})=VI$ color observed for each of the $N_*$ selected objects ($VI_{\rm obs}$) and that of the shifted D-MRL at the same level of magnitude ($VI_{\rm MRL}$), plus a second term where this color difference is weighted by taking into account both the photometric color error ($\sigma$) and the spatial distance ($d$) from the central star, according to the following equations:
\begin{equation}
  \Delta VI=\sum_{i=1}^{N_*}\left( \left | {VI_{{\rm obs},i} -VI_{{\rm MRL},i}} \right |
  + w_i \cdot \left | {VI_{{\rm obs},i} -VI_{{\rm MRL},i}} \right | \right)
\end{equation}
and 
\begin{equation}
w_i= \frac{1}{d_i\cdot \sigma_i} \left[{\sum_{j=1}^{N_*}\left(\frac{1}{d_j\cdot \sigma_j} \right)} \right]^{-1}
\end{equation}
The second  term is meant to give more importance to the closest stars (for a good spatial resolution of the resulting map), avoiding biases due to the use of a too large $r_{\rm lim}$. 
Note that we cannot reduce the equation (3) to the second term only, because a bright object (with small photometric errors) close to the central star would dominate the weight and make the reddening estimate essentially equal to the value needed to move the D-MRL exactly on that object.
We also performed a $3\sigma$ rejection in order to discard stars with a color distance from the D-MRL significantly larger than that of the bulk of the selected objects. 
This is useful to exclude field interlopers and/or discard non canonical stars \citep{pallanca10, pallanca13,  pallanca14, cadelano15a, cv6624, ferraro15, ferraro16a} with intrinsic colors different from those of the cluster main populations.
The final differential reddening value assigned to each star is the value of the $  {\rm E}(B-V)$ step that minimises the residual color ($\Delta VI$).

\section{DISCUSSION AND CONCLUSIONS}

We applied the procedure described in the previous section to each ``epoch/detector'' catalog  by following two different approaches. In the assumed magnitude range ($17.5<m_{\rm F814W}<21.5$) we considered:
(1) all the measured stars; (2) only the bona-fide cluster stars, selected for the {\it chi} and {\it sharpness} values that also survived to the proper motion selection.
Note that the first approach maximises the total spatial extension of the reddening map, since it allows to explore also the North-West portion of the FOV, which has been observed only in one epoch (EP4) and for which proper motions cannot be computed. 

\begin{figure*}
\begin{center}
\includegraphics[width=180mm]{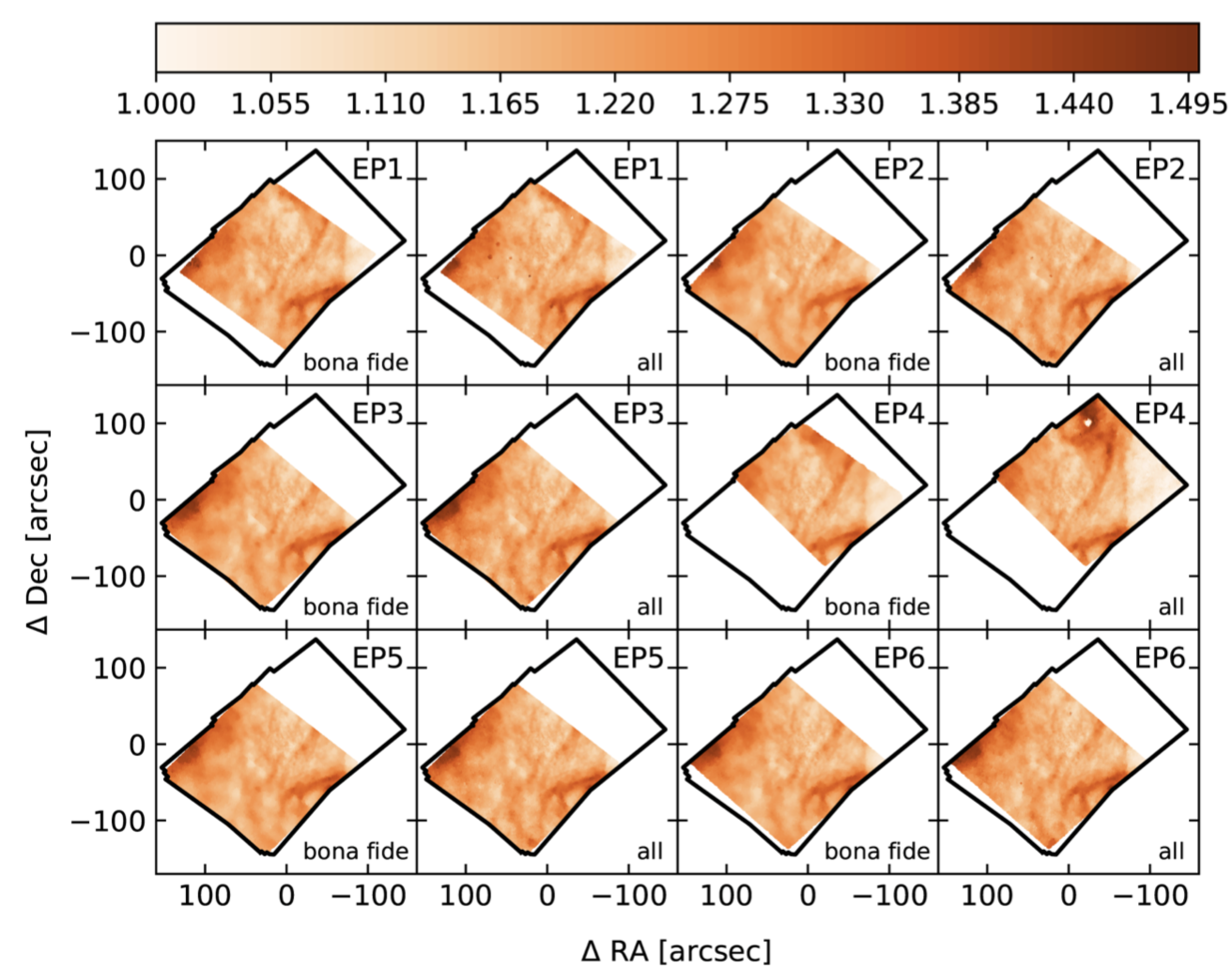}
\caption{ Reddening maps, centred onto the cluster center \citep{harris}, for the 6 different epochs 
and for the two approaches adopted here: by using only bona fide cluster members and all stars (see labels).    Thick black lines mark for clarity the contour of the total FOV. The top colorbar codifies the $ {\rm E}(B-V)$ values ranging from less extinct regions (white colors) up to more absorbed areas (darker colors).}\label{Fig:mapall}
\end{center}
\end{figure*}

In Figure \ref{Fig:mapall} we show the differential reddening maps obtained for all the six epochs with the two approaches. 
 These have been derived  directly from the values of $  {\rm E}(B-V)$ calculated per each single star,  visualising as a colormap the spatial behaviour of the reddening with a nominal resolution of a few $0.1 \arcsec$.
Obviously, each epoch covers  slightly different regions of the total FOV with significant overlaps. 
In  the panels sampling the same region of the sky, it is possible to recognise the same main structures. 
In particular a main feature is visible in all the panels: a sort of high-extinction filament crossing the South-West portion of the cluster. 
Other reddening filaments are visible crossing the cluster center approximately in the North-South direction.  
Moreover, from the direct comparison of the reddening values obtained, we found an extreme level of homogeneity among different epochs and between the two approaches. 
This is somehow expected because of the adopted method, but it testifies the correct application of the procedure and guarantees that the method is poorly affected by the presence of field interlopers.
 
\begin{figure*}
\begin{center}
\includegraphics[width=180mm]{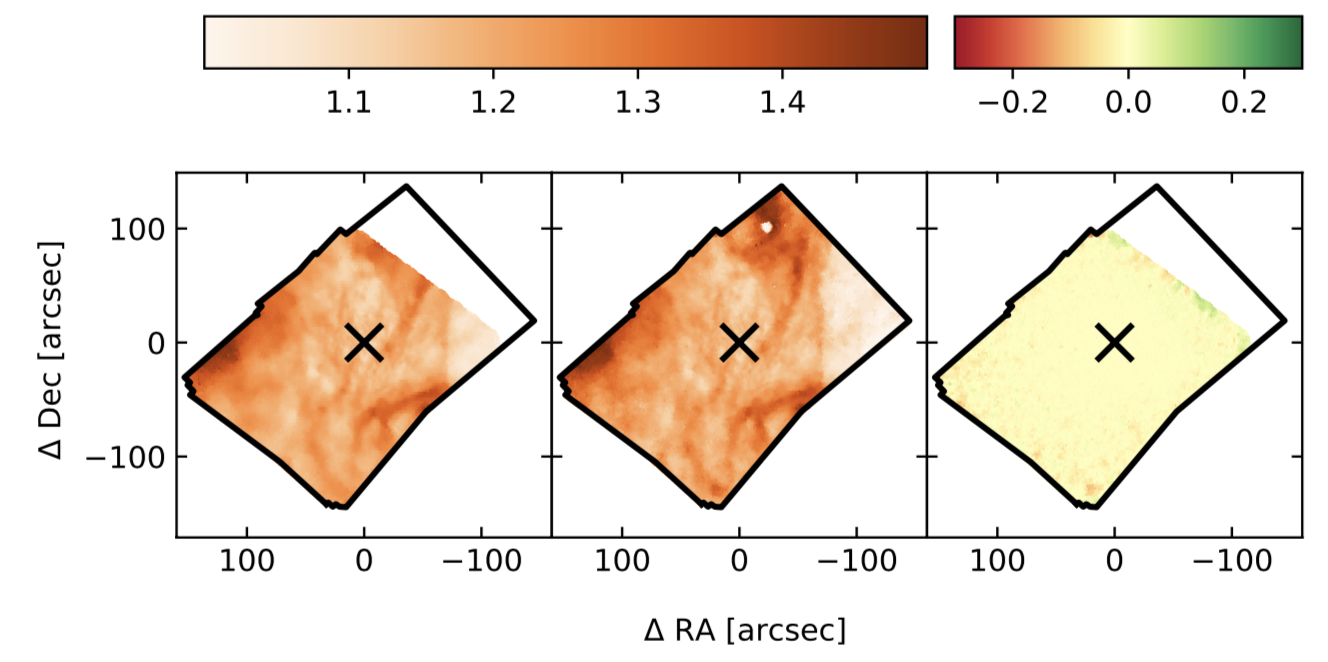}
  \caption{Global reddening maps computed adopting the two approaches described in the text: by using only bona fide cluster members ({\it left panel}) and all stars ({\it central panel}). The white spot located in the northern part of the FOV is due to the presence of a very bright (foreground) star that prevents the estimation of the local reddening. The residual map  resulting from the difference of the two maps in the overlapping region is shown in the {\it right panel}.  
 In each panel the center of the cluster is marked with a black cross, North is up and East is  on the left.    The thick black line marks for clarity the contour of the total FOV.  The top-left colorbar codifies the $ {\rm E}(B-V)$ values with respect to the D-MRL,  ranging from less extinct regions (lighter colors) up to more absorbed areas (darker colors). The top-right colorbar shows the colors assigned to the residual values.}
\label{Fig:diffmap}
\end{center}
\end{figure*}

To quantify the  agreement between the two approaches, we  directly compared the two corresponding global extinction maps. 
 To do that, we created a uniform grid of cells with side $0.5\arcsec$ and to each cell we assigned the average $  {\rm E}(B-V)$ calculated  as the median of all the available values (i.e. all the values assigned to all the stars within each cell in different epochs). 
Figure \ref{Fig:diffmap}  shows the extinction maps obtained for both approaches and the “residual maps”  resulting from their difference in the region in common. 
As can be seen the residual map is essentially flat, showing an average value lower than $ ~5 \%$, thus demonstrating that both the intensity and the spatial displacement of the main features are consistently estimated in the two approaches.
  
\begin{figure*}
\begin{center}
\includegraphics[width=180mm]{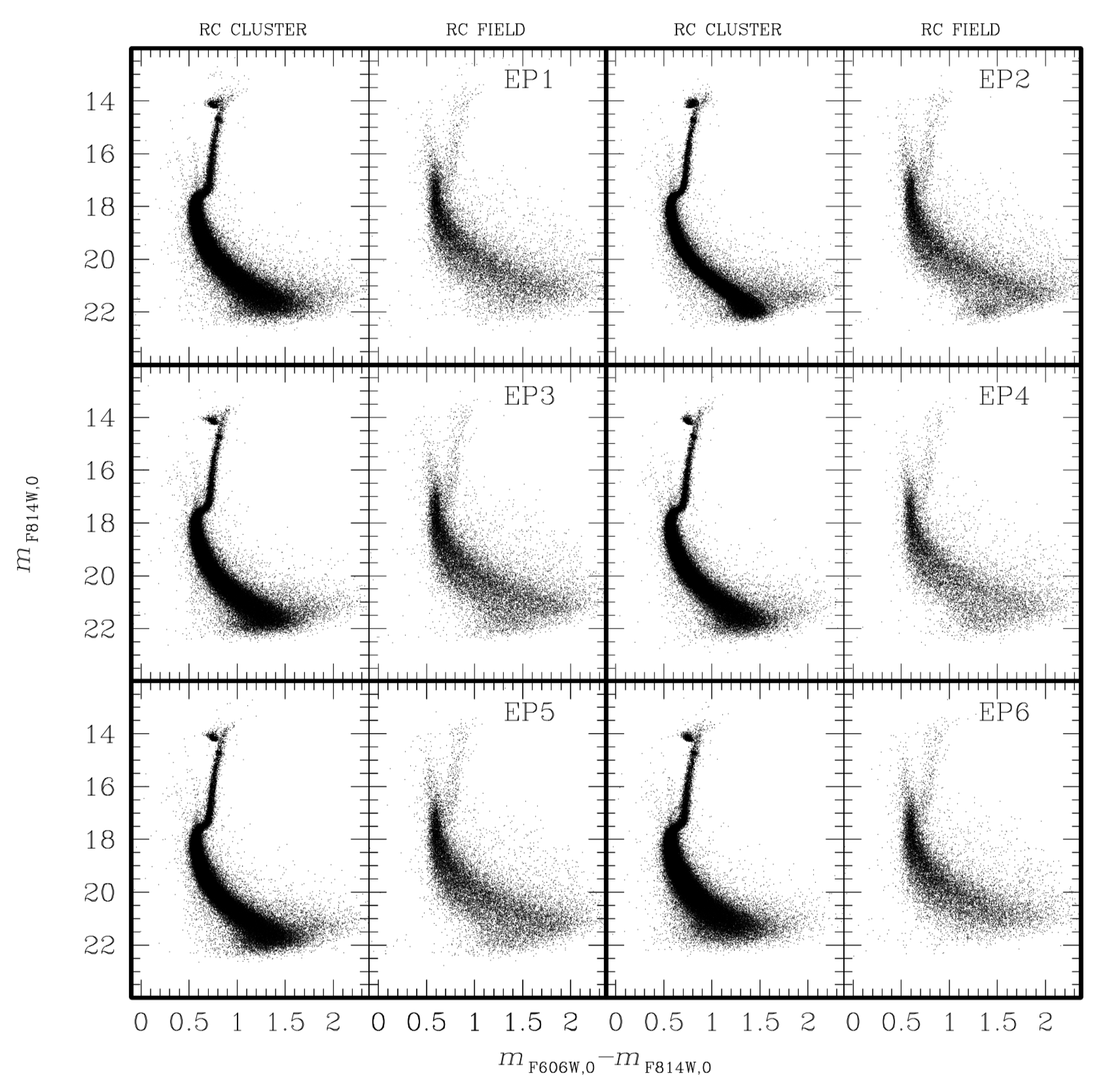}
  \caption{For each epoch (see labels) two reddening corrected (RC) CMDs are shown: in the left panel we plot all the candidate member stars, while the right panel shows the stars considered non members.}
\label{Fig:cmdDRC}
\end{center}
\end{figure*}

The differential reddening corrected CMDs, where the magnitudes and colors of each star have been corrected by using the values estimated through the ``bona-fide" approach, are shown in Figure \ref{Fig:cmdDRC}. 
Although the calculated reddening map is appropriate for cluster stars and not necessarily suitable for the field population, we show the effect of the correction on both the cluster and the field members. 
The comparison with Figure 3 highlights the clear improvement in the quality of the CMDs  obtained thanks to the the differential reddening correction: indeed all the main evolutionary sequences in the CMD are thinner and better defined. 
To quantify the effectiveness of the correction, we estimated the width of the cluster faint RGB (in the range $16.5<m_{\rm F814W,0}<15.5$) before and after the correction: the dispersion in color  decreases by more than a factor of 2, from   $= 0.05$ mag down to   $= 0.02$ mag.

The central panel of Figure \ref{Fig:diffmap}  shows the total differential reddening map in the direction of NGC 6440 computed in this study. 
It has a spatial resolution of only $0.5\arcsec$ and  ${\rm E}(B-V)$ spans $\sim 0.5$ mag, ranging from $\sim 1$  to $\sim1.5$ with a small tail at lower values. 
The most extincted region appears to be located in the northern portion of the cluster.  
Note that the white spot located in this region is an artifact due to a presence of a very bright foreground star that prevents the estimation of the local reddening. 
From this region a filamentary structure  extends toward the cluster center. 
Indeed the quite large spatial resolution of the map nicely shows other filamentary structures crossing the FOV, as that visible in the South-West portion of the cluster. 
Another clumpy cloud is visible toward  the East, while the West portion of the FOV shows a  substantial drop of the extinction approximately starting from $\sim 70\arcsec$ West of the cluster center, showing an essentially vertical (North-South direction) edge.  

\begin{figure*}
\begin{center}
\includegraphics[width=190mm]{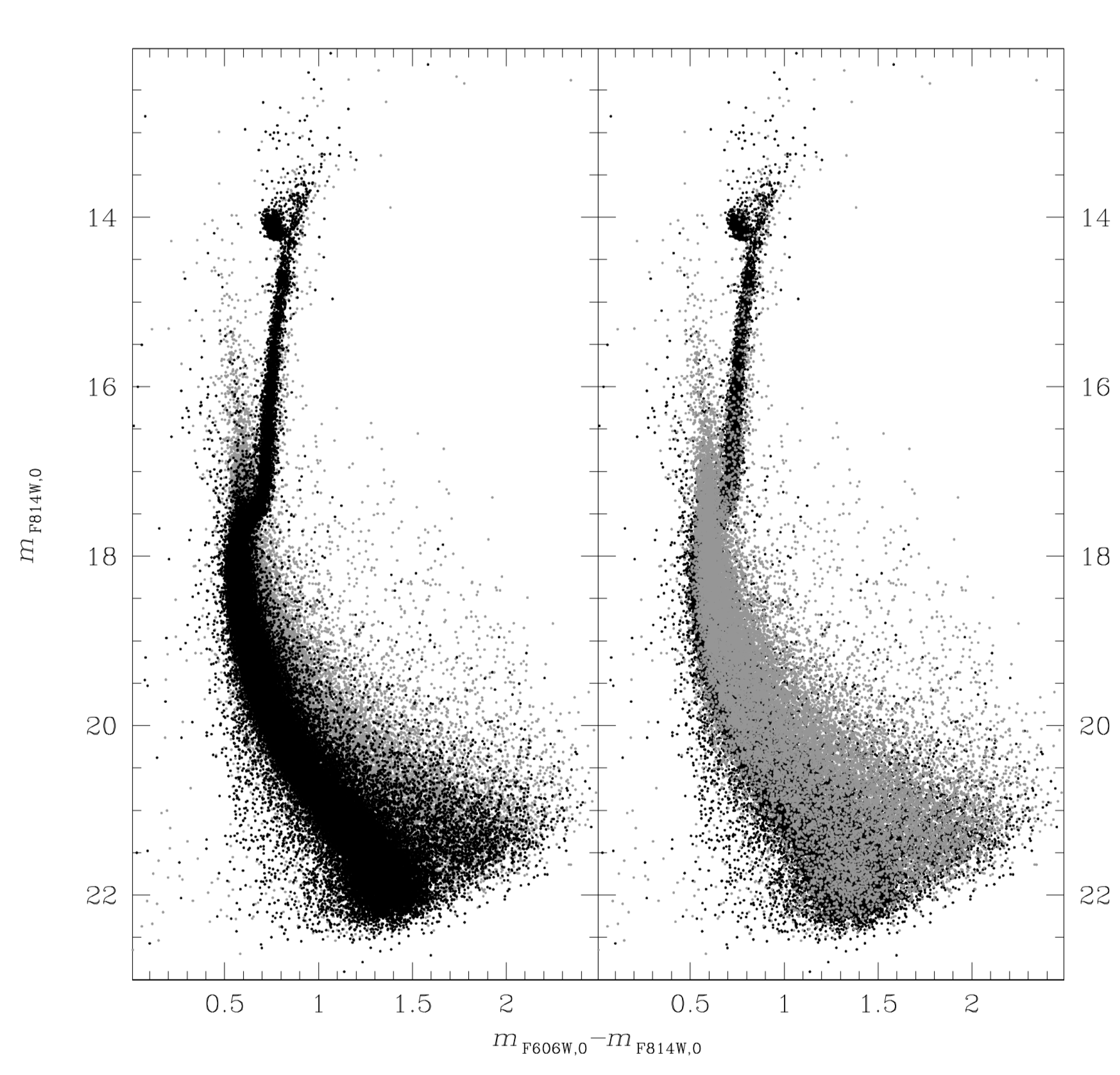}
  \caption{ The cleanest CMD of NGC 6440 obtained so far. The figure shows only stars selected for the {\it chi} and {\it sharpness} parameters, with the magnitudes corrected for the effect of differential reddening. Black and grey dots  correspond, respectively, to  stars which are likely cluster members and field interlopers (as determined from the proper motion analysis). To more clearly illustrate the CMD location of field stars with respect to the cluster population, the grey dots are plotted behind and in front of the black dots in the left and right panels, respectively.}
\label{Fig:totalcmd}
\end{center}
\end{figure*}

Finally,  Figure \ref{Fig:totalcmd} reports the differential reddening corrected and {\it chi} and {\it sharpness} cleaned CMD for both candidate member stars (113765 objects; black points) and  field stars (22873 stars; grey points). 
This is the cleanest and most accurate CMD obtained so far for NGC 6440.  
As can be seen, with the exception of the brightest portion of the RGB (saturated in these observations) all the evolutionary sequences are clearly defined.  
This dataset will be used in forthcoming studies aimed at deriving the physical structural parameters of the cluster (Pallanca et al, 2019, in preparation),  by using the method presented in \citep{miocchi13}, and to explore the internal dynamics of the core by means of complementary radial velocity measures (Ferraro et al, 2019, in preparation) acquired in the contest of the MIKiS  Survey \citep{ferraro18b,ferraro18c}. 

\acknowledgments
We thank the anonymous referee for the useful comments. This work is part of the project Cosmic-Lab at the Physics and Astronomy Department of the Bologna University. The research was funded by the MIUR throughout the PRIN-2017 grant awarded to the project {\it Light-on-Dark} (PI:Ferraro). The research is based on observations collected with the NASA/ESA HST (Prop. 11685, 12517, 13410 and 15403), obtained at the Space Telescope Science Institute, which is operated by AURA, Inc., under NASA contract NAS5-26555.  SS gratefully acknowledge funding from a European Research Council consolidator grant (ERC-CoG-646928-Multi-Pop).

\vspace{5mm}
\facilities{HST(WFC3/UVIS)};
\texttt{Matplotlib} (\citealt{matplotlibref}); 
\texttt{NumPy} (\citealt{numpyref}).



\end{document}